\documentstyle[12pt]{article}
\begin{document}
\begin{titlepage}
\begin{center}

\vskip 4em
  {\LARGE OPEN STRINGS WITH TOPOLOGICALLY} 
\vskip0.5cm   
  {\LARGE INSPIRED BOUNDARY CONDITIONS}
\vskip 5em
{\large Riccardo Capovilla${}^{\dagger}$ and Jemal Guven${}^{*}$}
\footnote{Permanent address:
Instituto de Ciencias Nucleares, 
 Universidad Nacional
Aut\'onoma de M\'exico, Apdo. Postal 70-543, 04510,
M\'exico, D.F., MEXICO}
\\[3em]
\em{
 ${}^{\dagger}$ Departamento de F\'{\i}sica \\
 Centro de Investigacion
y de Estudios Avanzados del I.P.N. \\
Apdo Postal 14-740, 07000 M\'exico,
D. F.,
MEXICO \\
}
\tt {capo@fis.cinvestav.mx} \\[1em]
\em{
$*$\ 
School of Theoretical Physics, 
D.I.A.S. \\
10 Burlington Rd., Dublin 4, IRELAND \\
 }
\tt{
 jemal@nuclecu.unam.mx
}
\\[1em]
\end{center}
\vskip 3em
\begin{abstract}
We consider an open string described by an action of the Dirac-Nambu-Goto 
type with topological corrections which affect the boundary conditions
but not the equations of motion. The most general addition of this 
kind is a sum of the Gauss-Bonnet action and the first Chern number 
(when the background spacetime dimension is four) of the normal bundle
to the string worldsheet. We examine the modification introduced 
by such terms in the boundary conditions at the ends of the string.  
\end{abstract}
\date{\today}

\end{titlepage}
\newpage


The simplest topological action for an open 
string which is dependent only on the intrinsic
geometry is proportional to the lorentzian version of
the Euler characteristic, $\chi$. Ignoring
boundary discontinuities, this is given by 

\begin{equation}
4\pi\chi  =    \int_m d^2\xi\sqrt{-\gamma}\,{\cal R}
+ 2\int_{\partial m} d\tau\, k\,.
\end{equation}
The bulk Gauss-Bonnet action must be supplemented with 
a boundary term to yield a topological invariant.
Here $\tau$ is the proper time,  
and $k$ the geodesic curvature along the end worldline.
Thus the introduction of the bulk Gauss-Bonnet type term into the action
for an open string
induces a load on the ends, and it is equivalent 
to attaching ``rigid" particles at the ends. 
The effect of adding this term to the Dirac-Nambu-Goto
action is to alter the boundary conditions at the ends,
leaving the equations of motion unaltered. This finds
an application in the ``stringy" description of hadronic
physics \cite{GSW,Polb,BN}.

At the same order, for a string living
in a four-dimensional background spacetime, there 
is an additional topological
invariant, constructed from the extrinsic geometry
of the string worldsheet, given by the first Chern
number of the normal bundle of the worldsheet.
This is related to the self-intersection number 
of the worldsheet. The addition of this term is
motivated by the construction of a string analogue of
the $\theta$-term in QCD \cite{Polb,Pol,Bal,MN}.

In this note, we present a simple derivation of
the appropriate boundary conditions for a Dirac-Nambu-Goto
open string  supplemented with such topologically
inspired terms in
the action. We use simple variational techniques, employing
the general results about
the variation of the intrinsic and
extrinsic geometry of a membrane of arbitrary
dimension in an arbitrary background previously derived
in Ref. \cite{CG1}. With respect to previous
treatments \cite{W1}, our approach has the advantage of
bringing to the forefront the geometrical 
content of the boundary conditions. Moreover,
we can easily derive the appropriate boundary
conditions when the background geometry is
arbitrary.  For additional recent studies on this
topic, see Refs. \cite{W2,HW,HR,KSW}.

The system is defined by the action functional

\begin{equation}
S[X] = - \mu I_0
- \alpha I_1 
- \beta   I_2 \,.
\label{eq:action}
\end{equation}
Here the field variable is $X$, the embedding functions
of the worldsheet $m$ swept out by the string in spacetime,
defined by $x^\mu = X^\mu (\xi^a )$, where
$x^\mu$ are local coordinates in the background spacetime
($\mu , \nu = 0,1,2,3$), and $\xi^a$ are coordinates
on the worldsheet $m$ ($a,b = 0,1$);
$\mu$ is the tension in the 
string, $\alpha$ and $\beta$ are two dimensionless numerical 
parameters. 

The first term is proportional to the area of
the worldsheet $m$, and is
known as the
Dirac-Nambu-Goto action, with the area defined by

\begin{equation}
I_0 = \int_m d^2\xi\sqrt{-\gamma}
\end{equation}
where $\gamma$ denotes the determinant of the metric induced
on $m$ by the background spacetime metric,

\begin{equation}
\gamma_{ab} = g_{\mu \nu} \partial_a X^\mu \partial_b X^\nu\,.
\end{equation}
The second term is proportional to the bulk Gauss-Bonnet
action,

\begin{equation}
I_1 = \int_m d^2\xi\sqrt{-\gamma}\, {\cal R} \,,
\end{equation}
where
${\cal R}$ is the scalar curvature constructed with 
the induced metric on the string worldsheet. 

The last term is defined by
\begin{equation}
I_2 =  \int_m d^2\xi\, \widetilde\Omega \,.
\end{equation}
The scalar density $\widetilde\Omega$ is constructed from the 
extrinsic twist curvature  of the extrinsic
twist potential, $\omega_a{}^{ij}$, the connection associated with the
freedom of normal rotations (see {\it e.g.} \cite{CG1})

\begin{equation}
\Omega_{ab}{}^{ij} = 
\partial_b \omega_a{}^{ij} 
- \partial_a \omega_b{}^{ij}\,,
\end{equation}
by contracting the normal ($i,j = 1,2$) and worldsheet index 
pairs with the corresponding Levi-Civita antisymmetric symbols:

\begin{equation}
\widetilde\Omega := {1 \over 2} \epsilon_{ij}\epsilon^{ab}
\Omega_{ab}{}^{ij}\,.
\end{equation}

It is now obvious that $\widetilde\Omega$ is a total divergence so that
$I_2$ is a topological invariant. In fact, it is proportional
to the first Chern number of the normal bundle of $m$.
We note that the tensor density $\epsilon^{ab}$ (assuming the values $0$ and 
$\pm 1$) is related to the tensor $\varepsilon^{ab}$
by $\epsilon^{ab} = \varepsilon^{ab}  \sqrt{-\gamma}$.

The curvatures ${\cal R}_{abcd}$ and $\Omega_{ab}{}^{ij}$ are 
related to the  
extrinsic and background geometry,
as characterized by the extrinsic curvature $K_{ab}{}^i$
and the spacetime Riemann tensor $R^{\mu}{}_{\nu\alpha\beta}$ 
by the following integrability conditions, the Gauss-Codazzi
and the Ricci equations, respectively: 

\begin{equation}
{\cal R}_{abcd}  = 
K_{ac}{}^i K_{bd\,i} - K_{ad}{}^i K_{bc\,i}
+ R_{\mu\nu\alpha\beta}\, e_a^\mu e_b^\nu e^\alpha_c e^\beta_d\,,
\label{eq:gc}
\end{equation}
and

\begin{equation}
\Omega_{ab}{}^{ij} =  
K_{ac}{}^i K_{b}{}^{c\,j}
 - K_{bc}{}^i K_{a}{}^{c\,j}
+ R_{\mu\nu\alpha\beta}\, e_a^\mu e_b^\nu n^{i\,\alpha} n^{j\,\beta} \,,
\label{eq:rc}
\end{equation}
where $e^\mu{}_a$ denote the two vectors tangent to the
worldsheet, and $n^{\mu\, i}$ the two unit vectors normal
to the worldsheet.

There is an additional constraint 
on the covariant derivative of the extrinsic 
curvature given by the 
the Codazzi-Mainardi equations:

\begin{equation}
\widetilde\nabla_a K_{bc}{}^i
- \widetilde\nabla_b K_{ac}{}^i
= R_{\mu\nu\alpha\beta} e_a^\mu e_b^\nu e_c^{\alpha} n^{i\,\beta} 
\,.
\label{eq:cm}
\end{equation}
Here $\widetilde\nabla_a$ denotes the $O(2)$ worldsheet covariant
derivative

\begin{equation}
\widetilde\nabla_a = \nabla_a - \omega_a\,.
\end{equation}
In particular, if the background geometry has constant curvature so 
that 

\begin{equation}
R_{\mu\nu\alpha\beta} = {R \over 12} (g_{\mu\alpha} g_{\nu\beta}-
g_{\mu\beta} g_{\nu\alpha})\,,
\end{equation}
 the projections appearing in both
Eqs.(\ref{eq:rc}) 
and (\ref{eq:cm}) vanish. It follows that 
$\Omega^{ij}{}_{ab}$ is completely constructed out of the extrinsic 
curvature, and it assumes the
simple form:

\begin{equation}
\widetilde\Omega =  \epsilon_{ij}\epsilon^{ab} K_{ac}{}^i K_{bd}{}^j
\gamma^{cd}\,.
\label{eq:omegaflat}
\end{equation}
The price one pays for this simple form, 
unfortunately, is that it is no longer obvious that
$\widetilde\Omega$ is related to a curvature. 

When the background spacetime
$M$ is flat, the worldsheet scalar
curvature
can also be expressed in terms of the extrinsic curvature, and
one has that

\begin{equation}
{\cal R} = K^i K_i - K_{ab}{}^i K^{ab}{}_i\,,
\label{eq:gausscon}
\end{equation}
where $K^i = \gamma^{ab} K_{ab}{}^i$ is the mean extrinsic curvature.

We are interested in the variation of the
functional $S[X]$ with respect to an infinitesimal 
variation of the embedding functions

\begin{equation} 
X^\mu \to X^\mu + \delta X^\mu\,.
\end{equation}
The variation can be decomposed into a part normal and a
part tangential to the worldsheet,

\begin{eqnarray}
\delta X^\mu &=& \delta_\perp X^\mu + \delta_\parallel X^\mu \nonumber \\ 
&=& \Phi_i n^{\mu\, i} + \Phi^a e^\mu{}_a\,.
\end{eqnarray}

For the variation of the Dirac-Nambu-Goto action, all that is needed is 
the well known variation of the intrinsic metric,

\begin{equation}
\delta \gamma_{ab} = 2 K_{ab}{}^i \Phi_i +
2 \nabla_{(a} \Phi_{b)}\,,
\label{eq:vargamma}
\end{equation}
where the second term is the worldsheet Lie derivative of $\gamma_{ab}$ 
along the worldsheet vector $\Phi^a$.
Using this expression, it is a simple exercise to obtain
the variation of the worldsheet area in the form,

\begin{eqnarray}
\delta I_0  &=&  \int_m d^2 \xi \sqrt{-\gamma}\,
[ K^i \Phi_i + \nabla_a \Phi^a ] \nonumber \\
&=&  \int_{m} d^2 \xi \sqrt{-\gamma} 
 K^i \Phi_i  + \int_{\partial m} d\tau \,\eta_a \Phi^a\,,
\label{eq:vardng}
\end{eqnarray}
where we have used Stoke's theorem; $\tau$ is the proper time 
induced on the string end by $\gamma_{ab}$,
and $\eta^a$ is the space-like unit normal to the boundary as embedded
in the worldsheet. 

The variation of the two topological terms by their nature
(we will show this explicitly) only
contributes in boundary terms. Only the Dirac-Nambu-Goto term makes a bulk 
contribution under a normal deformation of the 
string worldsheet. Therefore, the Euler-Lagrange equations for our system are  

\begin{equation}
K^i=0\,,
\label{eq:ki}
\end{equation}
the mean extrinsic curvature vanishes.
We comment  below on the additional boundary term in Eq.  
(\ref{eq:vardng}).

For the variations of the two topological terms it is 
useful to consider separately their tangential and normal variations.
For the former, we do not need to know anything about the 
specific form of the Lagrangian. For any
Lagrangian $L [X]$, since a tangential variation
is just a reparameterization of the worldsheet,we have that

\begin{equation}
\delta_{\parallel} \int_m d^2\xi\sqrt{-\gamma}\, L =
\int_{\partial m} d\tau  L 
\, \eta^a \, \Phi_a\,.
\end{equation} 
Hence

\begin{equation}
\delta_{\parallel} S[X] =
- \int_{\partial m} d\tau \left(\mu + \alpha {\cal R} + \beta 
\Omega \right) \eta^a\Phi_a\,,
\end{equation} 
where we have defined $\Omega$ with $\widetilde{\Omega}
= \Omega \sqrt{- \gamma}$.
Stationarity of the action under tangential deformations
requires that the Lagrangian 
must vanish on $\partial m$,

\begin{equation}
\mu + \alpha {\cal R} + \beta \Omega  = 0 \,.
\label{eq:b1}
\end{equation} 
This is a first boundary condition to be implemented
in addition to the equations of motion (\ref{eq:ki}).

It may appear that an alternative possibility is
to have the normal $\eta^a$ go null at the ends, as in the
standard boundary conditions for a Dirac-Nambu-Goto
open string, making the ends move at the speed of
light. However, since the introduction of $I_1$ implies
a load on the ends, as long as $\alpha \ne 0$,
generally these will follow a timelike
worldline, except at isolated points such as cusps.

Let us turn now to the normal variation of the
topological terms. For the sake of simplicity, we first assume 
that the background spacetime is flat. 
We will consider later an alternative strategy
when the background spacetime is arbitrary.

To obtain the normal variation of $I_1$, we 
use Eq. (\ref{eq:gausscon}) to express the scalar
curvature ${\cal R}$
in terms of the extrinsic curvature.
Under a normal displacement of the
worldsheet, when the background is flat,
the extrinsic curvature varies
according to \cite{CG1}:

\begin{equation}
\delta_\perp K_{ab}{}^i = - \widetilde\nabla_a \widetilde\nabla_b
\Phi^i + K_{ac}{}^i K_b{}^{c\,j} \Phi_j\,.
\label{eq:varK}
\end{equation}
Using  this expression,
together with Eq. (\ref{eq:vargamma}), one finds that

\begin{eqnarray}
\delta_\perp I_1 =  &2& \int_m d^2 \xi \sqrt{-\gamma}
\left( K^{ab}{}_i - K_i \gamma^{ab} \right) \widetilde\nabla_a
\widetilde\nabla_b \Phi^i \nonumber \\
= - &2&  \int_m d^2 \xi \sqrt{-\gamma} \left[ \widetilde\nabla_a
\left( K^{ab}{}_i - \gamma^{ab} K_i \right)\right] \widetilde\nabla_b
\Phi_i \nonumber \\
+ &2& \int_{\partial m} d\tau \eta_a 
\left( K^{ab}{}_i - K_i \gamma^{ab} \right) \widetilde\nabla_b \Phi^i
\,. 
\label{eq:var1}
\end{eqnarray} 
In the first line, there is a cancellation of terms that
do not involve derivatives of $\Phi^i$ which makes use of the
once-contracted Gauss-Codazzi equation
$ {\cal R}_{ab} = K_i K_{ab}{}^i - K_{ac}{}^i K_b{}^c{}_i $
and the two-dimensional fact ${\cal R}_{ab} = (1/2) {\cal R}
\gamma_{ab}$.
In the second line, we have integrated by parts, and we have
used Stoke's theorem in the second term.
The first integral over the worldsheet vanishes, because of
the contracted Codazzi-Mainardi equations (\ref{eq:cm}).

We need now to isolate the two independent variations on the boundary,
$\Phi^i$ and its derivative along the
normal to the boundary onto the worldsheet. 
To do this, it is convenient to exploit the completeness of the
orthonormal basis 
$\{\eta^a, v^a \}$ on the boundary, with $\eta^a$ the unit
vector normal to the boundary into $m$, and $v^a$ the unit vector
tangent to the boundary. Then, at the timelike boundary, we can write
the completeness relation

\begin{equation}
\gamma_{ab} = \eta_a\eta_b - v_a v_b\,,
\label{eq:gamma}
\end{equation}
and we can  decompose the worldsheet covariant derivative into
parts tangential and normal to the boundary as

\begin{equation}
\widetilde\nabla_a = \eta_a (\eta^b \widetilde\nabla_b) 
- v_a (v^b \widetilde\nabla_b ) =
\eta_a 
\widetilde\nabla_{(\eta)} 
- 
v_a \widetilde{\cal D}\,.
\label{eq:nabla}
\end{equation}
Here $\widetilde{\cal D} := v^a\widetilde\nabla_a =
\partial_\tau - v^a\omega_a^{ij}$.

Using these expressions in Eq. (\ref{eq:var1})
 we have that

\begin{equation}
\delta_\perp I_1 = 2 \int_{\partial m} d\tau
\left[ v^a v^b K_{ab}{}^i 
(\widetilde\nabla_{(\eta)} \Phi_i )
- \eta^a v^b K_{ab}{}^i 
( \widetilde{\cal D} \Phi_i ) \right]\,.
\end{equation}
We integrate by parts on the boundary to remove the derivative 
from $\widetilde{\cal D} \Phi_i$, and we define
the projections of the extrinsic curvature at the boundary
by
\begin{eqnarray}
K_{\parallel}{}^i &=& K_{ab}{}^i v^a v^b\,,\nonumber\\
K_{\parallel\perp}{}^i     &=& K_{ab}{}^i v^a \eta^b\,, \\
K_{\perp}{}^i         &=& K_{ab}{}^i \eta^a\eta^b\,, \nonumber
\end{eqnarray}
so that, finally, we obtain 

\begin{equation}
\delta_\perp I_1 = 2 \int_{\partial m} d\tau
\left[ K_{\parallel\,i} \left(\widetilde\nabla_{(\eta)} \Phi_i
\right) + \left(\widetilde{\cal D} K_{\perp\parallel}{}^i
\right) \Phi_i
\right]\,.
\end{equation}

Now, we evaluate the normal 
variation of  $I_2$, again
assuming that
the background is flat. We can use 
Eq.(\ref{eq:omegaflat}) to express 
$\widetilde\Omega$ in terms of the extrinsic
curvature, so that

\begin{eqnarray}
\delta_\perp I_2 &=& 2 \int_m d^2\xi \,
\epsilon_{ij}\epsilon^{ab} [ K_{a}{}^{d\,i}
( \delta_\perp K_{bd}{}^j ) + K_{ac}{}^i K^{a}{}_{d}{}^j (\delta_\perp
\gamma^{cd}) ]
\nonumber\\
&=& - 2 \int_m d^2\xi \,
\epsilon_{ij}\epsilon^{ab} K_a{}^{c\,i}
\widetilde\nabla_b\widetilde\nabla_c\Phi^j\,,
\end{eqnarray}
where we have used Eqs. (\ref{eq:vargamma}) and (\ref{eq:varK}). 

We now apply Stoke's theorem, to find  

\begin{equation}
\delta_\perp I_2 
= 2 \int_m d^2\xi \,
\epsilon_{ij}\epsilon^{ab} \widetilde\nabla_b K_{ac}{}^i
\widetilde\nabla^c\Phi^j
-
2 \int_{\partial m} d\tau \,
\epsilon_{ij}\varepsilon^{ab} \eta_b K_{ac}{}^i
\widetilde\nabla^c\Phi^j\,.
\end{equation}
The integral over $m$  
vanishes. To see this, we again 
exploit the Codazzi-Mainardi equations, Eq.(\ref{eq:cm}),
the right hand side of which is manifestly symmetric 
in $a$ and $b$.
This term therefore vanishes on contraction with $\varepsilon^{ab}$.
Finally, we again exploit Eq.(\ref{eq:nabla}) and 
Stoke's theorem on the 
end worldline to express $\delta_\perp I_2$ as a sum of two
independent variations:

\begin{equation}
\delta_\perp I_2 =
- 2 \int_{\partial m} d\tau \epsilon_{ij}
\varepsilon^{ab}\left[   
\widetilde{\cal D} 
(\eta_b K_{ac}{}^i v^c )\, \Phi^j
+ 
\eta_b \eta^c K_{ac}{}^i \,
\widetilde\nabla_{(\eta)} 
\Phi^j\right]\,..
\end{equation}
In the sequel it will be useful to exploit the identity

\begin{equation}
\varepsilon^{ab} = v^a \eta^b - \eta^a v^b\,.
\end{equation}
We note that

\begin{equation}
\varepsilon^{ab} v_b = \eta^a, 
\;\;\;\;\;\; \varepsilon^{ab}\eta_b = v^a\,.
\end{equation}
For the normal variation of $I_2$, we obtain  

\begin{equation}
\delta_\perp I_2 =
 2 \int_{\partial m} d\tau \epsilon_{ij}
\left[  
 K_{\perp\parallel}{}^j
(\widetilde\nabla_{(\eta)} 
\Phi^i ) + 
(\widetilde{\cal D} 
K_{\parallel}{}^j )\, \Phi^i \right]\,.
\end{equation}

As expected, the normal variation of the topologically
inspired terms gives only boundary terms involving
some projections of the extrinsic curvature along the
boundary , together with its first derivative along the
boundary.

It is now possible to write down the remaining
boundary conditions, in addition to Eq. (\ref{eq:b1}),
 for the system defined by (\ref{eq:action}) 
as 

\begin{equation}
 \widetilde{\cal D} \left[ \alpha K_{\parallel\perp}{}^i
+  \beta \epsilon^{ij}  K_{\parallel\,j} \right]
=0\,,
\label{eq:b2}
\end{equation}
and

\begin{equation}
\alpha K_{\parallel}{}^i + \beta \epsilon^{ij}
K_{\parallel\perp\,j} = 0\,.
\label{eq:b3}
\end{equation}

The boundary conditions, Eqs. (\ref{eq:b1}), (\ref{eq:b2}),
and (\ref{eq:b3}), supplement the bulk equations
of motion, Eq. (\ref{eq:ki}). Eqs. (\ref{eq:b1}) and
(\ref{eq:b3}) are of second order in time derivatives
of the ends embedding functions, whereas Eq. (\ref{eq:b2})
is of third order. 

We can now use the boundary conditions to evaluate
the curvatures at the boundary.
At the boundary, the curvatures are given by

\begin{eqnarray}
{\cal R} &=& 2(K_{\perp\parallel}{}^i K_{\perp\parallel\, i}
- K_{\parallel}{}^i  K_{\perp \, i})\,, \\
\Omega &=& 2 \epsilon_{ij} 
K_{\perp\parallel}{}^i (K_{\parallel}{}^j +
K_{\perp}{}^j )\,.
\end{eqnarray}

When $m$ is extremal, 
$K_{\perp}{}^i = K_{\parallel}{}^i $. This gives

\begin{eqnarray}
{\cal R} &=& 2
(K_{\perp\parallel}{}^i K_{\perp\parallel\,i}
- K_{\parallel}{}^i K_{\parallel\,i})\,, \\
\Omega &=& 4\epsilon_{ij} 
K_{\perp\parallel}{}^i  K_{\parallel}{}^j\,.
\end{eqnarray}
On the boundary, using this together with
(\ref{eq:b3}), one obtains 

\begin{eqnarray}
{\cal R} &=&  {2\over\beta^2} (\alpha^2 - \beta^2)
K_{\parallel}{}^i K_{\parallel\,i}\,, \\
\Omega   &=& 4{\alpha\over\beta}K_{\parallel}{}^i K_{\parallel\, i}\,.
\end{eqnarray}
The boundary condition (\ref{eq:b1}) then fixes the squared norm, 
$K_{\parallel}{}^i K_{\parallel\, i}$, completely in terms of the three 
parameters, $\mu,\alpha$ and $\beta$:

\begin{equation}
K_{\parallel}{}^i K_{\parallel\, i}
=  {\mu \beta^2\over 2\alpha} {1\over \alpha^2 +\beta^2}
 \,.
\label{eq:K2}
\end{equation}
This in turn fixes the boundary values of the curvatures:

\begin{eqnarray}
{\cal R} &=& {\mu\over \alpha} \left({\alpha^2 -\beta^2
\over \alpha^2 + \beta^2}\right)\nonumber\\
\Omega   &=& {2\mu\beta\over \alpha^2 + \beta^2}\,.
\end{eqnarray}
We note, in particular, that when $\beta\ne 0$,
$\alpha>0$ is necessary for the consistency of
Eq.(\ref{eq:K2}). $\beta$ may, however, assume either sign.

We have seen that the norm of $K_{\parallel}{}^i$ is fixed 
completely on the ends by the parameters 
appearing in the action, $\mu$, 
$\alpha$ and $\beta$. The remaining boundary condition Eq.(\ref{eq:b2})
modulo (\ref{eq:b3}) implies that

\begin{equation}
\widetilde{\cal D} K_{\parallel}{}^i = 0\,.
\end{equation}
$K_{\parallel}{}^i$ is covariantly 
constant along the end. Note that this is consistent with the
fact that the norm is fixed. The freedom is to choose the 
ratio, $K_{\parallel}{}^2 /K_{\parallel}{}^1$.
This we input as an initial condition. 

Of course, the worldsheet cannot be totally geodesic,
$K_{ab}{}^i = 0$, since this would imply $\mu = 0$.
Moreover, no solutions to our system exist in which 
the string moves on a plane, such that the extrinsic
curvature along one normal is vanishing, say $K_{ab}{}^1 = 0$,
for the same reason. In order to allow for planar solutions,
we must set $\beta = 0$.

These boundary conditions may also be seen in terms 
of the geometric quantities associated with the direct
embedding of the ends worldlines in spacetime \cite{CGSt}.
Then $K_\parallel{}^i $ coincides with the extrinsic
curvature of the end worldline along the unit normal vectors
$n^{\mu\,i}$. The extrinsic twist potential of the ends is
made up by $K_{\parallel\perp}{}^i $, which is the mixed part
along $n^{\mu\,i}$ and $\eta^{\mu}$, 
and by $v^a \omega_a{}^{ij}$.  The only geometric
quantity of the ends that does not appear in the
boundary conditions is $k$, the extrinsic
curvature along $\eta^a$.

It is interesting to consider various reduction of the system
defined by $S$. 
When $\alpha = 0$, we can no longer assume
that the trajectory will be timelike, and (\ref{eq:b1}) 
is substituted by the requirement that the ends go null. 
The boundary condition (\ref{eq:b3})
then implies that $\Omega$ vanishes, so that the extrinsic
twist potential is pure gauge at the ends.  
Also when $\beta = 0$, we have that 
$\Omega$ vanishes at the ends; moreover, now ${\cal R}$ is proportional to
the norm of $K_{\perp\parallel}{}^i$, and it is positive 
definite.

Let us suppose that in addition massive particles,
of mass $M$, are attached to 
the ends. Using the results of Refs. \cite{B1,B2}, we have that 
the boundary conditions (\ref{eq:b1}) and (\ref{eq:b3})
are modified respectively to

\begin{equation}
\mu + \alpha {\cal R} + \beta 
\Omega + M k  = 0 \,,
\end{equation}
where $k$ denotes the geodesic curvature of the ends into
$m$, and

\begin{equation}
(\alpha +  M/2) K_{\parallel}{}^i + \beta \epsilon^{ij}
K_{\parallel\perp\,j} = 0\,.
\end{equation}
Eq.(\ref{eq:b2}) is unchanged.
As before, $K_{\parallel}{}^i $ is covariantly constant
which in turn implies that its norm is too. This means that
$k$ is also constant on the end. The end particle moves with 
constant acceleration exactly as it would without the topological
modification.

We consider now the case in which the background spacetime geometry
is left arbitrary. In this case there is no calculational 
advantage in employing the integrability conditions
(\ref{eq:gc}) and (\ref{eq:rc}). 
We follow an alternative strategy, which also
provides an independent check of the validity of the
boundary conditions we have derived above.
 
To obtain the normal variation of $I_1$,
we need to know how
the scalar curvature varies. One has that,

\begin{equation}
\delta {\cal R}^a{}_{bcd} = 
\nabla_c (\delta \gamma_{db}{}^a )
- \nabla_d (\delta \gamma_{cb}{}^a )\,,
\end{equation} 
where $\gamma_{ab}{}^c$ is the affine connection
compatible with $\gamma_{ab}$.
For the scalar curvature, ${\cal R} = {\cal R}^a{}_{bad} \gamma^{bd}$,
we have

\begin{equation}
\delta_\perp{\cal R} =
\nabla_c (\gamma^{ab}\delta_\perp \gamma_{ab}{}^c)
- \nabla^b (\delta_\perp \gamma_{ab}{}^a ) 
- 2 {\cal R}_{ab} K^{ab}_i \Phi^i\,,
\end{equation}
where \cite{CG1}

\begin{equation}
\delta_\perp \gamma_{ab}{}^c =
\gamma^{cd}\left[ \nabla_a (K_{bd}{}^i \Phi_i) + \nabla_b (K_{ad}{}^i \Phi_i)
-\nabla_d (K_{ab}{}^i \Phi_i)\right]\,.
\end{equation}
We therefore obtain

\begin{equation}
\delta_\perp I_1 =
2\int_{\partial m} d\tau\, \eta^b\left[
\nabla_a (K^a{}_b{}^i\Phi_i)
- \nabla_b (K^i\Phi_i) 
\right]\,,
\label{eq:varr}
\end{equation}
where we have used the 2-dimensional identity
\begin{equation}
{\cal R}_{ab} = {1 \over 2} \gamma_{ab} {\cal R}\,,
\end{equation}
to reduce the variation to a total divergence, and Stoke's theorem
to express it as an integral over the ends.

To express Eq. (\ref{eq:varr}) in terms of the independent variations
at the boundary, we use Eqs. (\ref{eq:gamma}) and 
(\ref{eq:nabla}), so that

\begin{eqnarray}
\delta_\perp I_1 &=&
2 \int_{\partial m} d\tau \Big[\left( \eta^a\eta^b \widetilde\nabla_{(\eta)} 
K_{ab}{}^i - \widetilde\nabla_{(\eta)} K^i 
- v^a \eta^b \widetilde{\cal D} K_{ab}{}^i
+ \widetilde{\cal D} (v^a\eta^b K_{ab}{}^i)\right)\Phi_i\nonumber\\
&& + ( \eta^a\eta^b 
 K_{ab}{}^i
- K_i )
\widetilde\nabla_{(\eta)} 
\Phi_i\Big]\,.
\label{eq:vari1}
\end{eqnarray}
We note that the sum of the first two terms gives

\begin{eqnarray}
\eta^a\eta^b \widetilde\nabla_{(\eta)} 
K_{ab}{}^i
- \widetilde\nabla_{(\eta)} K^i 
&=&  v^a v^b \eta^c \widetilde\nabla_c K_{ab}{}^i\nonumber\\
&=&  v^a v^b \eta^c\widetilde\nabla_b K_{ac}{}^i
+ R_{\mu\nu\alpha\beta} \,v^\mu \eta^\nu v^{\alpha} n^{\beta\, i} \nonumber\\
&=&  v^a \eta^b \widetilde{\cal D} K_{ab}{}^i
+ R_{\mu\nu\alpha\beta}\, v^\mu \eta^\nu v^{\alpha} n^{\beta\, i}\quad,
\label{eq:gca}
\end{eqnarray}
where we have exploited the Codazzi-Mainardi equations
(\ref{eq:cm}) on the second line, and 
$v^\mu = e^\mu{}_a v^a\,, \eta^\mu = e^\mu{}_a \eta^a $. 
As a result, the first three terms 
on the right hand side of Eq.(\ref{eq:vari1}) get replaced by the 
projection of the spacetime Riemann tensor appearing in 
(\ref{eq:gca}).
Hence, we find that the normal variation of $I_1$ is now
given by

\begin{equation}
\delta_\perp I_1 =
 2\int_{\partial m} d\tau \left[ 
 K^i_{\parallel} (\widetilde\nabla_{(\eta)}\Phi_i )
+ (\widetilde{\cal D} K_{\parallel\perp}{}^i  +
R_{\mu\nu\alpha\beta}\, v^\mu \eta^\nu v^{\alpha} n^{\beta\, i})\Phi_i
\right]\,.
\end{equation}

For the normal variation of $I_2$, we use the fact that
the extrinsic twist curvature varies according to

\begin{equation}
\delta ( \Omega_{ab}{}^{ij} \epsilon_{ij} )  = 
\nabla_b (\delta\omega_a{}^{ij} \epsilon_{ij} ) -
\nabla_a (\delta\omega_b{}^{ij} \epsilon_{ij} )\,.
\end{equation}
In particular, under a normal deformation of the 
string worldsheet \cite{CG1},

\begin{equation}
\delta_\perp (\omega^{ij}_a \epsilon_{ij} ) =
- 2 K_{ab}{}^i\widetilde\nabla^b\Phi^j \epsilon_{ij} 
+ R_{\mu\nu\alpha\beta}\, n^{\mu\, i} n^{\nu\, j} n^{\alpha\, k}
e^\beta{}_a \Phi_k \epsilon_{ij} 
\,.
\end{equation}
We have then 

\begin{eqnarray}
\delta_\perp I_2 &=&  \int_{\partial m} d\tau\,\epsilon_{ij}
\varepsilon^{ab}\eta_b
(- 2 K_{ac}{}^i \widetilde\nabla^c \Phi^j +
 R_{\mu\nu\alpha\beta}\, n^{\mu\, i} n^{\nu\, j} n^{\alpha\, k} 
e^\beta{}_b \Phi_k)\nonumber\\
 &=&  
 \int_{\partial m} d\tau\,
\left[  
2 (\widetilde{\cal D}K^j_{\parallel})\epsilon_{ij}\Phi^i
+ R_{\mu\nu\alpha\beta}\, n^{\mu\,k} n^{\nu\, l} n^{\alpha\, i} 
v^\beta\Phi_i \epsilon_{kl}
+ 2 K_{\perp\parallel}{}^j \epsilon_{ij}
(\widetilde\nabla_{(\eta)} \Phi^i )\right]\,.
\end{eqnarray}

Therefore we find that the boundary conditions
(\ref{eq:b1}) and (\ref{eq:b3}) are not affected by the generalization
to an arbitrary background spacetime, whereas
(\ref{eq:b2}) is modified to

\begin{equation}
 \widetilde{\cal D} \left[ \alpha K_{\parallel\perp}{}^i
+ \alpha R_{\mu\nu\rho\sigma} v^\mu \eta^\nu v^\rho n^{\sigma\,i}
+  \beta \epsilon^{ij}  K_{\parallel\,j} 
+ {\beta \over 2} R_{\mu\nu\rho\sigma} n^{\mu\,k}
n^{\nu\,l} n^{\rho\,i} v^\sigma \epsilon_{kl} \right]
=0\,,
\label{eq:b2c}
\end{equation}
In particular, note that the additional
curvature terms vanish when the background spacetime
has constant curvature.

\vspace{1cm}
 
\noindent{\bf Acknowledgements}

\vspace{.5cm}

We received partial support from CONACyT under Grant
211085-5-0118PE.

\newpage

\end{document}